\documentclass[prb,nofootinbib,superscriptaddress,showpacs,floatfix]{revtex4}
\usepackage{bbding}
\usepackage{mathrsfs}
\usepackage{amsmath,amsfonts,amssymb}
\usepackage{epsfig}
\usepackage{graphicx}
\usepackage{wasysym}
\usepackage{bbm}
\usepackage{psfrag}
\usepackage{color}
\usepackage{pstool}
\usepackage{braket}
\usepackage{lmodern}
\usepackage[dvips, bookmarks, colorlinks=true, plainpages = false, citecolor = blue, linkcolor = blue, urlcolor = blue, filecolor = blue]{hyperref}
\def\be{\begin{equation}}
\def\ee{\end{equation}}
\def\bea{\begin{eqnarray}}
\def\eea{\end{eqnarray}}

\begin{document}
\title{Fidelity susceptibility of Su-Schrieffer-Heeger model with
  further neighbour hopping term}
\author{Surajit Mandal}\email{surajitmandalju@gmail.com}
\author{Asim Kumar Ghosh}
 \email{asimkumar96@yahoo.com}
\affiliation {Department of Physics, Jadavpur University, 
188 Raja Subodh Chandra Mallik Road, Kolkata 700032, India}
\begin{abstract} In this study, topological phase transition in
Su-Schrieffer-Heeger (SSH) model  with a 
further neighbour hopping term has been studied in terms of
fidelity susceptibility. Topological phase transition point
in the standard SSH model has been identified before by noting the
divergence of fidelity susceptibility. 
The same approach has been
employed here where fidelity susceptibility
is found to diverge at the phase transition points.
Additionally, effect of staggered potential
on the fidelity susceptibility has been explored.
Analytic expression for fidelity susceptibility has been obtained
along with its numerical estimation on finite chains 
by exact diagonalization. Fidelity susceptibility
exhibits sharp peaks at the phase transition points in both approaches.
Scaling exponent of this divergence has been
obtained numerically which is found to agree to that of
the standard SSH model without further neighbour terms.
\end{abstract}
\maketitle
\section{INTRODUCTION}
Quantum phase transitions (QPTs) are qualitative changes
in the ground-state properties of a quantum many-body system
driven by quantum fluctuations at zero temperature
\cite{Sachdev,Sondhi,Vojta,Hertz,Continentino}. Unlike
classical phase transitions, which are induced by thermal
fluctuations at non-zero temperatures,
QPTs occur when a non-thermal control
parameter in the Hamiltonian is varied across a
critical point. At the critical points, the
ground-state wavefunction undergoes a significant
restructuring, making quantum information-theoretic
measures valuable tools for characterizing critical behavior \cite{NielsenChuang,Amico2008}.
Among these measures, fidelity has emerged
as a powerful and model-independent probe of
quantum criticality \cite{Zanardi2006,You2007,Gu2010}.
Fidelity quantifies the
similarity between two quantum states and is
defined as the overlap between ground states
corresponding to slightly different values of
a control parameter. For a Hamiltonian $H(\lambda)$
depending on a parameter $\lambda$, the ground-state fidelity is given by
\begin{equation}
F(\lambda,\delta)=
\left|
\langle \psi_0(\lambda)|
\psi_0(\lambda+\delta)
\rangle
\right|,
\label{FS}
\end{equation}
where $|\psi_0(\lambda)\rangle$ denotes the ground state
of the system and $\delta$ accounts a small increment of the
driving parameter.
However, fidelity can be defined analogously for any arbitrary states,
$|\psi(\lambda)\rangle$, besides the
ground state, $|\psi_0(\lambda)\rangle$. 
The fidelity takes values between $0$ and $1$,
where $F=1$ indicates identical states and
$F=0$ corresponds to orthogonal states.
Obviously, fidelity satisfies the relation, $F(\lambda,0)=1$,
in Eq. \ref{FS}, for the normalized ground state of the system.
Near a quantum critical point, the
ground state changes rapidly with the variation of driving parameter,
leading to a pronounced drop in fidelity.
Although fidelity is not itself
an entanglement measure, it provides
valuable information about the global
structure of quantum states. In
many-body quantum systems, a sudden decrease
in fidelity signals a significant change
in the ground-state wavefunction, which
often accompanies quantum phase transitions
driven by changes in entanglement.

\subsection{Fidelity Susceptibility as Measures of Quantum Correlations}
To quantify the sensitivity of the fidelity in a parameter-independent manner,
the concept of fidelity susceptibility is
introduced. Expanding the fidelity for an
infinitesimal parameter variation, $\delta\rightarrow 0$ yields
\begin{equation}
F(\lambda,\delta)
\simeq
1-\frac{1}{2}\chi_{\rm F}(\lambda)\,\delta^2,
\end{equation}
where the fidelity susceptibility  $\chi_{\rm F}(\lambda)$ is
\be
\chi_{\rm F}(\lambda)=-\frac{\partial^2F}{\partial \delta^2}.
\ee
It measures the leading response of the ground state to
a small perturbation and provides information about the
geometry of the quantum state manifold. Importantly,
fidelity susceptibility often exhibits singular
behavior or diverges at quantum critical points,
thereby serving as a sensitive indicator of phase transitions.
Fidelity susceptibility quantifies the sensitivity
of the ground state to infinitesimal
variations of the driving parameter.
Near a quantum critical point,
the ground-state wavefunction changes
rapidly, leading to a pronounced peak or
even divergence of $\chi_F$ in the thermodynamic limit.
Since quantum phase transitions
originate from qualitative changes in quantum
correlations and entanglement, fidelity susceptibility
serves as an indirect measure of entanglement and critical
behavior. It has been successfully applied to spin chains,
Bose-Hubbard systems, topological phases, and other strongly
correlated quantum systems. Compared with conventional
entanglement measures, fidelity susceptibility is
computationally advantageous because it depends
only on the overlap of neighboring ground states of the whole system 
rather than the explicit evaluation of reduced density matrices.

The study of fidelity susceptibility has
attracted considerable interest because
it does not require prior knowledge of
an order parameter or symmetry-breaking
mechanism. Consequently, it can be applied
to a wide range of systems, including
topological phases, strongly correlated
quantum systems, and models exhibiting
unconventional quantum criticality.
Furthermore, its connection to the
quantum geometric tensor establishes a
deep relationship between quantum information
theory and condensed matter physics \cite{Venuti2007,Quan2006}.
As a result, fidelity susceptibility has
become an essential tool for investigating
critical phenomena, finite-size scaling
behavior, and the universal properties of
quantum phase transitions \cite{Sachdev}. Its ability to
reveal subtle changes in the structure of
quantum states makes it a valuable quantity
in the study of modern quantum many-body systems.

In the next section utility of  fidelity susceptibility
will be explored in the context of topological phase transition in
the Su-Schrieffer-Heeger (SSH) model.
The analytic expression of fidelity
susceptibility will be derived in the momentum space and the
character of divergence of fidelity susceptibility 
at the transition point will be obtained in terms of
algebraic power-law behaviour. 
In parallel, fidelity susceptibility
will be estimated numerically in the real space.
Both approaches lead to convergence in the thermodynamic limit.
In this context, SSH chains with odd and even number of sites will be
considered. At the same time, effect of staggered and uniform
chemical potential will be discussed. 
Subsequently, properties of fidelity susceptibility
for extended SSH model with the presence of
single further neighbour hopping terms will be
presented and its power-law behaviour will be studied at the transition points. 
\section{THE Su-Schrieffer-Heeger (SSH) model}
\label{SSH}
The SSH model is a one-dimensional
tight-binding model that describes electrons hopping on a
chain with alternating hopping amplitudes\cite{Qi,SSH1,SSH2,SSH3}. Originally
introduced to explain the electronic properties of
polyacetylene, it has become a fundamental model for
studying topological phases of matter in one-dimension.

\begin{figure}[h]
\psfrag{A}{\large A}
\psfrag{B}{\large B}
\psfrag{t}{\large $t$}
\psfrag{t1}{\large $t'$}
\includegraphics[width=300pt]{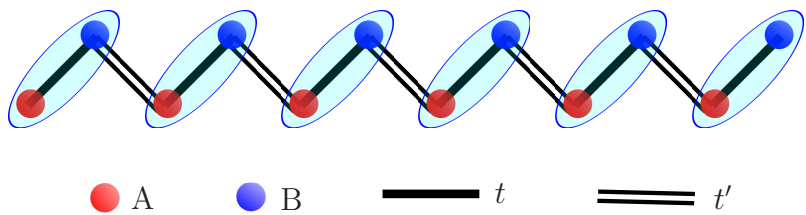}
\caption{Geometry of Su-Schrieffer-Heeger (SSH) model. 
  Two sites marked by A and B constitute unit cell as shown
  within shaded region.}
 \label{SSH-fig}
\end{figure}
Structure of SSH model has been depicted in Fig. \ref{SSH-fig},
whose dimer unit cell has been shown by shaded region. 
Within a unit cell, two nonequivalent sites have been labelled by 
A and B, connected by a bond with hopping amplitude $t$, 
while adjacent unit cells are connected by another bond
of different hopping amplitude $t'$. 
The dimer unit cell accommodates two nonequivalent sites, A and B.

The tight-binding (TB) Hamiltonian for the SSH model is
given by,
\be H_{\rm {SSH}}=\sum_{j=1}^{N/2} \left( t \,a_j^\dag b_j+t'\,b_j^\dag a_{j+1}
+h.c.\right),
 \label{SSH-ham}
 \ee
where $a_j$,  and $b_j$, are the fermionic annihilation
operators corresponding to the A and B sites, 
respectively, of the $j$ th cell,
while $N$ is the total number of sites.

In order to diagonalize the Hamiltonian
(Eq. \ref{SSH-ham}), following 
Fourier transformation of the operators is considered: 
\[\gamma_j=\frac{1}{\sqrt N}\sum_{k\in {\rm BZ}} e^{ikj}\gamma_k,\]
where $\gamma_j=a_j$, or, $b_j$, and $k$ is the Bloch wave vectors  
within the Brillouin zone (BZ). 
Hamiltonian in the $k$-space 
can be written in terms of a $2\times 2$ matrix
for the dimer unit cell as:  
\be H_{\rm {SSH}}=\sum_{k\in{\rm {BZ}}}\Psi_k^\dag H(k)\Psi_k.\ee
The state vector is $\Psi_k=[a_k\;b_k]^{\rm T}$, 
where T stands for transpose,  
and
  \be H(k)=\mathbf h \cdot {\boldsymbol \sigma},\label{SSH-k}
  \ee
  where by assuming the lattice parameter as unity, 
         \[\left\{\begin{array}{l}h_x( k)=t+t'\cos{( k)}, \\[0.3em]
      h_y( k)=t'\sin{( k)}, \\[0.3em]
       h_z( k)=0,
         \end{array}\right. \]
         and the Pauli matrices are 
           \[\sigma_x = \left(\begin{array}{cc}0 &\;1\\
    1&\;0\end{array} \right)\!,\;
      \sigma_y= \left(\begin{array}{cr}0 &- i\\
        i&\,0\end{array} \right)\!,\;
      \sigma_z= \left(\begin{array}{cr}1 &0\\
        0&-1\end{array} \right)\!. \]
Now choosing new fermionic operators, $\alpha_k$ and $\beta_k$ for the
Bogoliubov transformation as,
\be
 \left(\begin{array}{c}\alpha_k\\
   \beta_k \end{array} \right)= \frac{1}{\sqrt 2}
 \left(\begin{array}{cc}g &\;1\\
    -g&\;1\end{array} \right) \left(\begin{array}{c}a_k\\
   b_k \end{array} \right),
    \ee
    where $g=e^{i\phi}$, $\phi = \arctan{\{t'\sin{k}/(t+t'\cos{k})\}}$,  
    the diagonalized Hamiltonian can be written as
    \be H_{\rm {SSH}}=\sum_{k\in{\rm {BZ}}}
    \left(\alpha_k^\dag\; \beta_k^\dag\right)
    \left(\begin{array}{cc}\epsilon_k &\;0\\
    0&\;-\epsilon_k\end{array} \right) \left(\begin{array}{c}\alpha_k\\
      \beta_k \end{array} \right).\ee
Hence the ground state of the system at half-filling limit is obtained by
filling up all the negative energy states ($-\epsilon_k$), and when all the
positive energy states, $\epsilon_k$, are empty. As a result, the
ground state can be expressed as
\be |\Psi_0\rangle = \prod_{k\in{\rm {BZ}}}\beta_k^\dag|0\rangle. \label{GS}\ee
We know that the SSH model exhibits two distinct phases. 
When $t' > t$, it exhibits a topological
insulating phase which is characterized by the non-zero winding number,
$\nu=1$, and the emergence of
zero-energy edge states under open
boundary conditions. When $t' < t$,
the system enters in a trivial insulating phase because of the fact that
in this case $\nu=0$. The topological
transition occurs at $t=t'$, where the bulk energy gap closes and
band inversion takes place.
\subsection{Fidelity susceptibility for the SSH model}
\subsubsection{Open boundary condition:}
Fidelity susceptibility in this case can be defined by the Eq. \ref{FS},
or, $F(\lambda,\delta)=\left|\langle \psi_0(\lambda)|\psi_0(\lambda+\delta)
\rangle\right|,$
where $\lambda = t/t'$, and the ground state, $\psi_0$, be the Slater
determinant constructed out of the $N/2$ lower energy states
under the open boundary condition in the half-filled limit, when $N$
is even. It corresponds to the fact that one of the edge state 
contributes into the ground state when the system is topologically nontrivial. 
On the other hand, for odd $N$, ground state in terms of
Slater determinant can be constructed out of
either $(N-1)/2$ or $(N+1)/2$ particles into the system.
It means that the zero energy edge mode is either empty or occupied.
Since the Hamiltonian is particle-hole symmetric,
the fidelity is, however, exactly the same in both cases.
It is expected that the overlap between the two many-particle
wave functions, $\left|\langle \psi_0(\lambda)|
\psi_0(\lambda+\delta)\rangle\right|$,
tend to vanish exponentially with the
system size, $N$, when $\delta \ne 0$, which is similar to
the Anderson orthogonality catastrophe \cite{Anderson}.
It is therefore suggested to define the fidelity susceptibility
in terms of the fidelity density as \cite{Sirker},
\be
f(\lambda,\delta)=-\frac{1}{N} \,\ln{F(\lambda,\delta)}. 
\ee
The value of $f(\lambda,\delta)$ is not only zero when $\delta =0$,
since $F(\lambda,0)=1$, but it is the minimum, so,
\[\left.\frac{\partial f}{\partial \delta}\right|_{\delta=0}=0.\]
Hence under the Taylor series expansion:
\[f(\lambda,\delta)=\chi_{ f}(\lambda)\,\delta^2 +\mathcal O (\delta^3),\]
   where the fidelity susceptibility in this case is
   \[\chi_{ f}(\lambda)=\frac{1}{2}\left.\frac{\partial^2 f}
         {\partial \delta^2}\right|_{\delta=0}.\]
At the phase transition point, $\chi_{f}(\lambda)$ 
exhibits universal scaling behavior irrespective of system size
and thus can be used to characterize the phase transition.
Thus it is pertinent to discuss the behaviour of $\chi_{f}(\lambda)$
in the vicinity of topological phase transition point of the SSH model.
As the SSH model essentially represents a many-body fermionic system
composed of non-interacting electrons, the ground state,
$\psi_0$, can be expressed by a simple Slater determinant
of the single particle eigenstates. The fidelity $F(\lambda,\delta)$
in this case can be written as
\be F(\lambda,\delta)=\left| {\rm Det} P(\lambda,\delta)\right| \ee
where $P$ is a $M\times M$ overlap matrix whose elements
can be expressed by the single particle states as
\be P_{lm} (\lambda,\delta)
=\langle \psi_l(\lambda)|\psi_m(\lambda+\delta) \rangle, \ee
where $M$ is the number of particles in the ground state.
The fidelity susceptibility in this formalism can be expressed as\cite{Sirker}:
\be
\chi_{ f}(\lambda)=-\frac{1}{N}
\lim_{\delta\rightarrow 0}\frac{\ln{F(\lambda,\delta)}}{\delta^2}.
\label{chi-N}
\ee
Value of $\chi_{ f}(\lambda)$ thus can be estimated by
obtaining the eigenvectors numerically by diagonalizing the
Hamiltonian matrix under open boundary condition separately for
even and odd number of sites and under periodic boundary condition
for even number of sites. Obviously, for even number of sites $M=N/2$,
while for odd number of sites $M=(N\pm 1)/2$, as stated earlier.

\begin{figure}[h]
  \psfrag{N}{$N$}
  \psfrag{O}{Odd $N$}
   \psfrag{E}{Even $N$}
\psfrag{c}{$\chi_{ f}(\lambda)$}
\includegraphics[width=300pt]{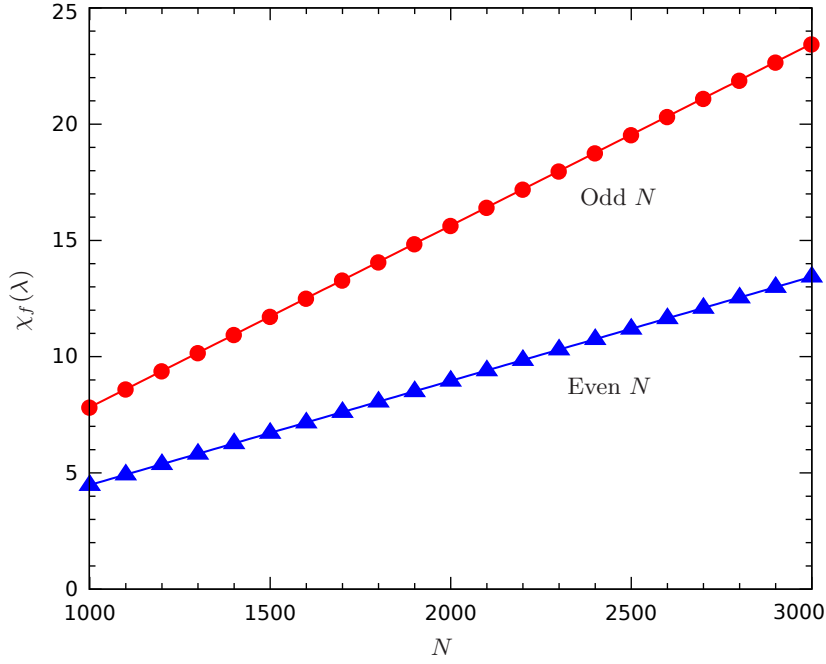}
\caption{Finite size scaling of $\chi_{ f}(\lambda)$ at the
  phase transition point ($t/t'=1$) for odd $N$ (red circles) and even $N$ (blue triangles).
  Straight line fitting are obtained with $\chi_{ f}(\lambda) \approx
  0.00782 N$, for odd $N$ and  $\chi_{ f}(\lambda) \approx
  0.00448 N$, for even $N$.}
 \label{line-fitting}
\end{figure}
In order to demonstrate the finite size effects of the
numerical data, $\chi_{ f}(\lambda)$ at the phase transition point, $t/t'=1$,
for different values of odd and even $N$ have been plotted in Figure
\ref{line-fitting}. 
Numerical data are found to fit with the straight lines
with different slopes for odd and even numbers $N$.
For odd $N$, data fit with $\chi_{ f}(\lambda) \approx
0.00782 N$,  and  for even $N$ data fit with
$\chi_{ f}(\lambda) \approx 0.00448 N$. Although
$\chi_{ f}(\lambda) \propto N$, for both the cases, 
slope for the odd $N$ is found higher that that of even $N$.
However, the slopes of these straight lines indicate that
$\chi_{ f}(\lambda)$ will diverge in the thermodynamic limit
($N\rightarrow \infty$), in both cases. 

\begin{figure}[h]
    \psfrag{a}{(a)}
  \psfrag{b}{(b)} 
  \psfrag{E}{Energies}
  \psfrag{o}{$N=199$ (Odd)}
    \psfrag{e}{$N=200$ (Even)}
   \psfrag{0.0}{0.0}
\psfrag{t}{$t/t'$}
\includegraphics[width=300pt]{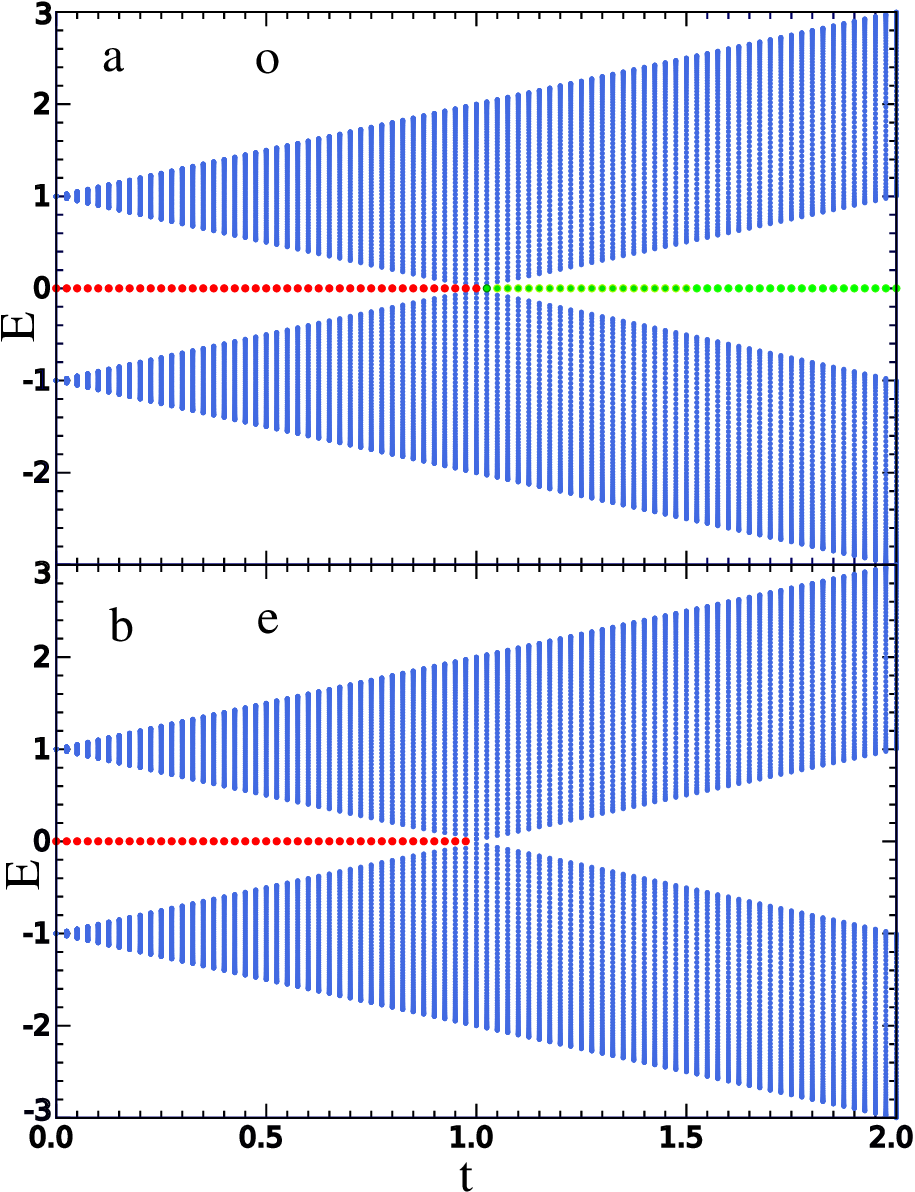}
\caption{Energy spectrum for open SSH chain with respect to $t/t'$
  for odd (a) and even (b) number of $N$. Zero energy edge modes are
  shown in red and green spheres.}
 \label{energy-spectrum}
\end{figure}

In order to understand the difference in slopes for odd and even
values of $N$, energy spectrum of the SSH model for odd and even cases
are shown in Figure \ref{energy-spectrum} (a) and (b), respectively.
For even $N$, (Figure \ref{energy-spectrum} (b)),
a pair of zero-energy edge modes appear
in the topological regime, $t/t'<1$, which are shown in red circles.
This pair of edge states are in accordance to the
bulk-boundary correspondence rule for the topological
phase with winding number $\nu=1$. 
These edge modes are found to emerge in each of the two edges or
on both the sublattices, A and B. This feature is consistent with
the global chiral symmetry of the system. In the trivial regime, $t/t'>1$, no
zero energy edge mode appears. On the other hand, for odd N,
only a single zero energy edge mode appears althrough the trivial and nontrivial
topological regimes. Edge modes in the trivial region 
$t/t'>1$ (non-trivial region $t/t'<1$ ) are shown in green (red) 
circles in Figure \ref{energy-spectrum} (a).
However, band gap vanishes at the phase
transition point $t/t'=1$, for both the cases as observed in Figure 
\ref{energy-spectrum} (a) and (b).

\begin{figure}[h]
    \psfrag{a}{(a)}
  \psfrag{b}{(b)} 
  \psfrag{E}{Energies}
  \psfrag{N}{$N=199$}
  \psfrag{d1}{$t/t'=1/2$}
    \psfrag{d2}{$t/t'=2$}
  \psfrag{p1}{\color{red} $|\psi_{100}|^2$}
   \psfrag{p2}{\color{green} $|\psi_{100}|^2$}
\psfrag{s}{sites}
\includegraphics[width=300pt]{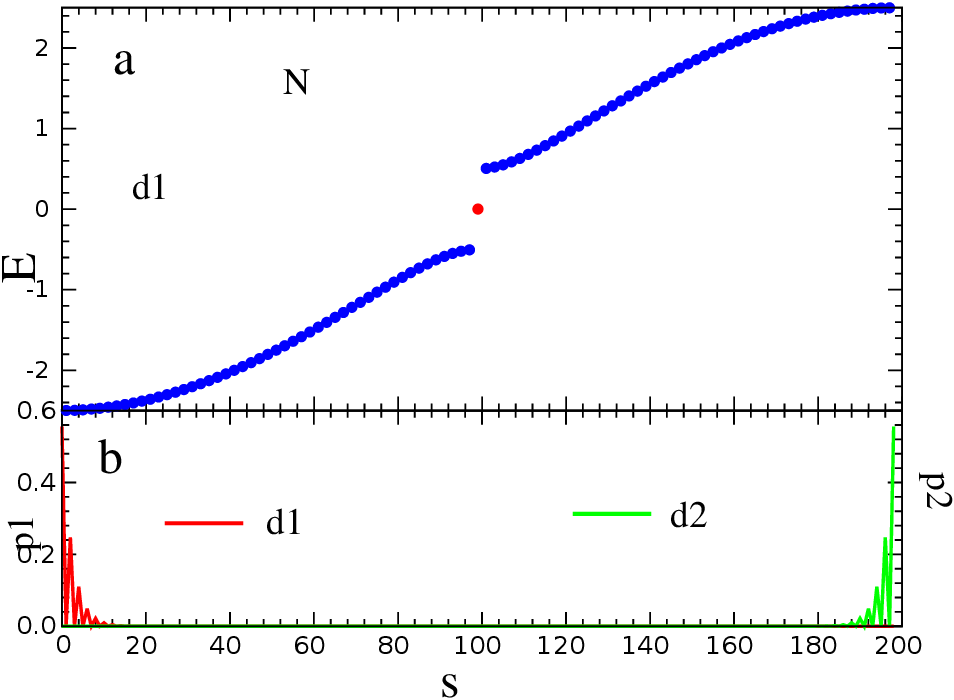}
\caption{(a) Energy spectrum for open SSH chain 
  for odd number of sites, $N=199$. Zero-energy edge mode is 
  shown in red sphere. (b) Amplitude density of edge mode, $|\psi_{100}|^2$,  
  in the trivial (red line) and nontrivial (green line) phases.}
 \label{energy-amplitude-odd-N}
\end{figure}

Energy spectrum for open SSH chain with odd number of sites
when $N=199$ has been shown
in Figure \ref{energy-amplitude-odd-N} (a), where a single zero-energy edge
mode is found in both trivial ($t/t'>1$) and nontrivial regions ($t/t'<1$). 
Probability density of the edge mode,  
$|\psi_{100}|^2$ per site in the nontrivial phase has been
plotted in Figure \ref{energy-amplitude-odd-N} (b)
in red line when $t/t'=1/2$. In this case
probability density is found nonzero (zero) in the A (B) sublattice.
Probability density of the edge mode, $|\psi_{100}|^2$ per site
in the trivial phase has been shown when $t/t'=2$. Again,
probability density is found nonzero (zero) in the A (B) sublattice
like the nontrivial case as shown in Figure \ref{energy-amplitude-odd-N} (b)
in green line. It is worth mentioning to note that zero-energy edge mode 
found in odd-$N$ system no more obey the conventional bulk-boundary correspondence rule.
According to bulk-boundary correspondence rule a pair of zero-energy edge states
must be found in the open chain when the winding number is $\nu =1$ in the nontrivial phase
when number of sites, $N$ is even.
On contrary, for odd-$N$ only one edge mode is found which contradicts the
conventional bulk-boundary correspondence rule. It means the edge of the system behaves
in different way for the odd-$N$ and even-$N$ models as far as the topological
property is concerned.
It is expected that this dissimilarity attributes to the difference in slopes
for odd- and even-$N$ cases as observed in Figure \ref{line-fitting}. 

In addition, variations of $\chi_{ f}(\lambda)$ with respect to $t/t'$ for $N=1001$
(odd-$N$) and for $N=1000$ (even-$N$) have been shown in Figure
\ref{chi-even-odd} (a) and (b), respectively with red circles.
Sharp peaks have been observed at the phase trasition point
$t/t'=1$, both for odd-$N$ and even-$N$ cases which
are found to agree with the analytic results. 
\subsubsection{Periodic boundary condition:}
Apart from the numerical estimation of the value of
fidelity susceptibility for even $N$ under both open and periodic boundary
conditions, analytic expression of $\chi_{ f}(\lambda)$ for the SSH model can be
obtained for even $N$ under periodic boundary condition.
Using the ground state vector obtained in Eq. \ref{GS}
and the definition of fidelity by the Eq. \ref{FS},
expression of $\chi_{ f}(\lambda)$ can be obtained.
At the half-filling limit the overlap integral can be simplified as
 \bea F(\lambda,\delta)&=&\left|\langle \Psi_0(\lambda)|\Psi_0(\lambda+\delta)
 \rangle\right|,\nonumber \\[0.3 em]
 &=& \prod_{k,k'}\left|\langle0|\beta_k\, \beta_{k'}^\dag|0\rangle\right|,\nonumber \\
 &=& \prod_{k}\left[\frac{1}{2}\bigg(1+g_k(\lambda)\, g^*_{k}(\lambda+\delta) \bigg) \right].
 \eea
 Henceforth, the fidelity density can be obtained in a straight
 forward manner\cite{Sirker}.
 \bea
 f(\lambda,\delta)&=&-\frac{1}{N} \,\ln{F(\lambda,\delta)},\nonumber \\[0.3 em]
 &=&-\frac{1}{N} \sum_k \ln{\left[\frac{1}{2}\bigg(1+g_k(\lambda)\, g^*_{k}(\lambda+\delta) \bigg) \right]},\nonumber \\[0.3 em]
  &=&-\frac{1}{2\pi}\int_{-\pi}^\pi dk \,\ln{\left[\frac{1}{2}\bigg(1+g_k(\lambda)\, g^*_{k}(\lambda+\delta) \bigg) \right]},
 \eea
 where $g_k(\lambda)=e^{i\phi}$, $\phi = \arctan{\{t'\sin{k}/(t+t'\cos{k})\}}$,
 while $g_k(\lambda+\delta)=e^{i\tilde \phi}$,
 $\tilde\phi = \arctan{\{t'\sin{k}/((t+\delta) +t'\cos{k})\}}$.
Expression for fidelity susceptibilty therefore obtained as 
\bea \chi_{ f}(t,t')&=&\frac{1}{2}\left.\frac{\partial^2 f}
     {\partial \delta^2}\right|_{\delta=0},\nonumber \\[0.3 em]
       &=&\frac{1}{4\pi} \int_{0}^{\pi} dk \left(\left.\frac{\partial \theta}{\partial t} \right|_{\delta=0}\right)^2,
      \eea
     where $\theta=(\phi-\tilde \phi)$. Upon further simplification,
     \be \chi_{ f}(t,t') =\frac{1}{16\,(t+t')}\times
     \left\{\begin{array}{l}
     \frac{1}{(t-t')},\;{\rm when},\;t>t',\\[0.4em]
      \frac{t^2}{t'^2\,(t'-t)},\;{\rm when},\;t<t'.
  \end{array}\right.
     \label{chi}\ee
It reveals that $\chi_{ f}(t,t')$ diverges at the phase transition point
$t=t'$, according to the algebraic power-law, 
\be
\chi_{ f} = \frac{A}{|t-t'|^\beta},
\label{scaling-SSH}
\ee
where $A$ and $\beta$ are the scaling amplitude and exponent, respectively,  
with $\beta=1$. This feature is consistent
with the general scaling theory at the phase transition point which
establishes that $\chi_{ f}\sim |t-t'|^{d\eta-2}$, where
$d$ is the spatial dimension of the system or $d=1$,
and $\eta$ is the critical exponent related to
the divergence of the correlation length at the critical point.
Since $d\eta-2=-1$, it estimates that exponent of divergence of correlation
length at the topological phase transition point in this case is $\eta=1$ \cite{Sirker}.

\begin{figure}[h]
  \psfrag{a}{(a)}
  \psfrag{b}{(b)} 
   \psfrag{t}{$t/t'$}
\psfrag{c}{$\chi_{ f}(\lambda)$}
\includegraphics[width=330pt]{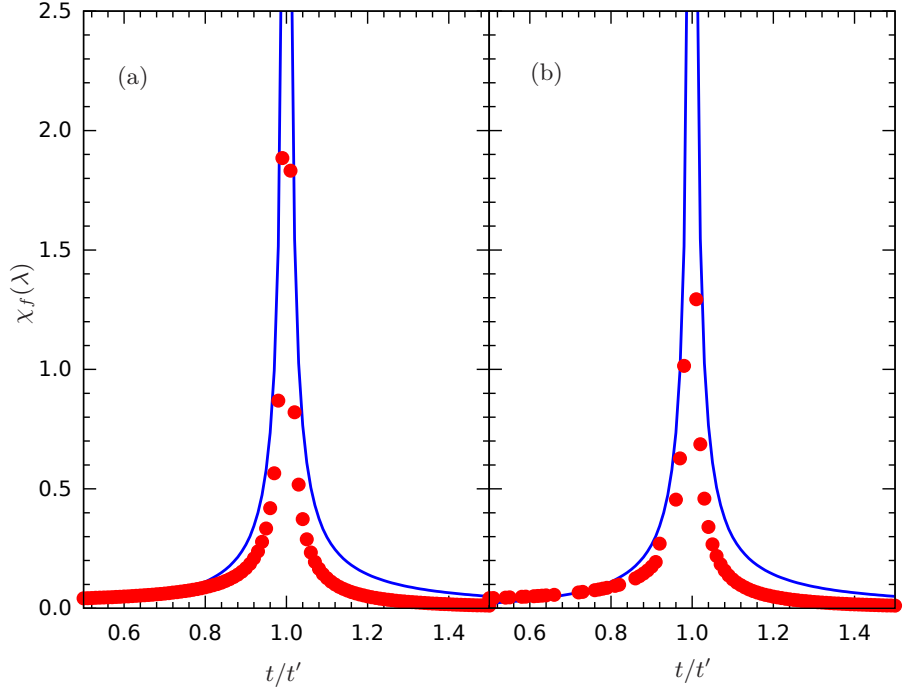}
\caption{Variation of $\chi_{ f}(\lambda)$  with respect to
  $t/t'$ in red circle for $N=1001$ (odd $N$) in (a)
  and $N=1000$ (even $N$) in (b).
  Analytic result as given in Eq. \ref{chi} is drawn in blue line
  for both in (a) and (b). }
 \label{chi-even-odd}
\end{figure}

Variation of $\chi_{ f}(\lambda)$  with respect to  $t/t'$
has been shown in Figure \ref{chi-even-odd} (a) and (b) in blue line.
Analytic results (Eq. \ref{chi}), has been compared with the
numerical values for both $N$ odd ($N=1001$) and even ($N=1000$)
as shown in (a) and (b), respectively. Eq. \ref{chi} has been derived for the
fidelity susceptibility of the bulk system, as the periodic boundary condition
has been imposed. On contrary, numerical values are
obtained for open boundary conditions. So, they contain the contributions both from 
bulk and boundary of the system. Closer agreement between analytic and
numerical values can be found if contribution from the boundaries are
taken off by suitable scaling of the numerical data
for the open chains \cite{Sirker}.

Another notable feature of the $\chi_{ f}(\lambda)$ is the asymmetry of
the variation around the phase transition point $t/t'=1$, which is
found both in analytic and numerical results in Figure \ref{chi-even-odd}.
This feature can be explained
by noting that the system for $t/t'=0$ in the left most side of
the phase transition point ($t/t'=1$) actually represents the extreme case,
where it is a collection of isolated dimers in the topologically
nontrivial phase. This particular point is not shown in
Figure \ref{chi-even-odd}. On the other hand, in the right side of the
phase transition point another extreme point $t'/t=0$ will be approachable when
 $t\rightarrow \infty$, which is far away from the
phase transition point. In this extreme case, the system
represents again a collection of disjoint dimers but in the topologically
trivial phase. However, perfect symmetry of the $\chi_{ f}(\lambda)$ around
the phase transition point $t/t'=1$ can be achieved by suitable scaling of the
parameters\cite{Sirker}.
\subsection{SSH model in staggered potential}
\begin{figure}[h]
  \begin{center}
\psfrag{zt}{$Z_{\rm t}$}
\psfrag{zb}{$Z_{\rm b}$}
\psfrag{z}{$Z$}
  \psfrag{t}{ $t/t'$}
  \psfrag{mu}{\hskip 0.05 cm $\mu=0.1$}
 \includegraphics[width=290pt]{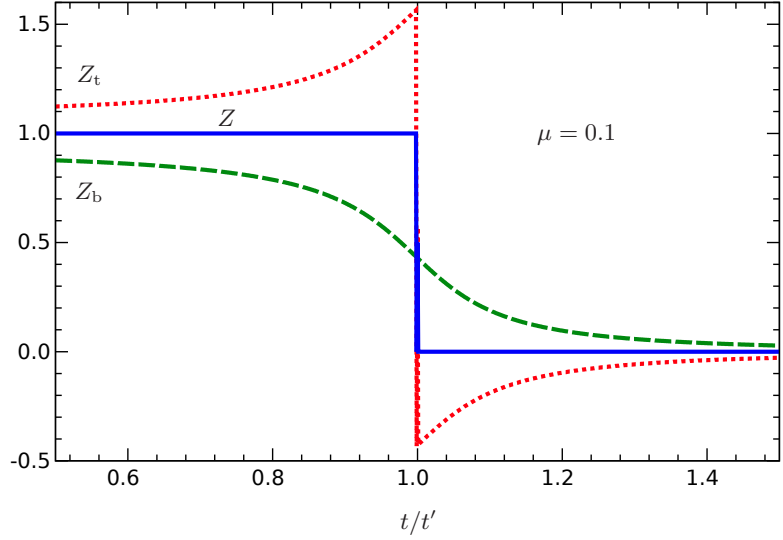}
 \caption{Variation of Zak numbers, $Z_{\rm t}$, $Z_{\rm b}$, 
   and average Zak number, 
   $Z=(Z_{\rm t}+Z_{\rm b})/2$, with $t/t'$ for the SSH model.}
\label{zak-ssh}
\end{center}
\end{figure}

The Hamiltonian for SSH model in the presence of staggered potential is given by
\be H_{\rm {SSH\text{-}SP}}=\sum_{j=1}^N \left[ \left( t \,a_j^\dag b_j+t'\,b_j^\dag a_{j+1}
+h.c.\right) + \mu \left(a_j^\dag a_j-a_j^\dag a_j\right) \right],
 \label{SSHSP-ham}
 \ee
 where $\mu$ be the chemical potential. Hamiltonian in the momentum space can be
 expressed as 
 \be H(k)=\mathbf h' \cdot {\boldsymbol \sigma},\nonumber
  \ee
  where 
         \[\left\{\begin{array}{l}h'_x( k)=t+t'\cos{( k)}, \\[0.3em]
      h'_y( k)=t'\sin{( k)}, \\[0.3em]
       h'_z( k)=\mu,
         \end{array}\right. \]
In contrast to the SSH model, Hamiltonian in Eq. \ref{SSHSP-ham}
breaks the chiral symmetry. As a result, winding number cannot be used to study
its topological character.
Instead, Zak number ($Z_n$) which is nothing but Zak phase per $\pi$, or  
\be
Z_n=\frac{i}{\pi}\int_{-\pi}^\pi dk \langle \psi_n(k)|\partial_k| \psi_n(k)\rangle,
\nonumber
\ee
can be employed to determine the topological character of the $n$-th
energy band. 
Zak phase is basically the Pancharatnam-Berry phase evaluated 
for systems of one spatial dimension.
Zak numbers of the top ($Z_{\rm t}$, red dotted line)
and the bottom ($Z_{\rm b}$, green dashed line) bands along with 
the average Zak number, $Z=(Z_{\rm t}+Z_{\rm b})/2$, (blue line)
with varying $t/t'$ have been shown in Figure
\ref{zak-ssh}, when $\mu=0.1$. Although $Z_{\rm b}$ varies smoothly with $t/t'$,
surprisingly, $Z_{\rm t}$ encounters a sudden jump when $t/t'=1$. And as a result,
average Zak number also experiences a sudden jump at the point, $t/t'=1$.
Even though this feature reminds the topological phase transition at the
point $t/t'=1$, in terms of the Zak number $Z$, but the SSH model exhibits
no change of topology in the presence of staggered potential. 
\begin{figure}[h]
  \begin{center}
      \psfrag{E}{Energies}
  \psfrag{t}{ $t/t'$}
  \psfrag{mu}{\hskip 0.05 cm $\mu=0.1$}
 \includegraphics[width=290pt]{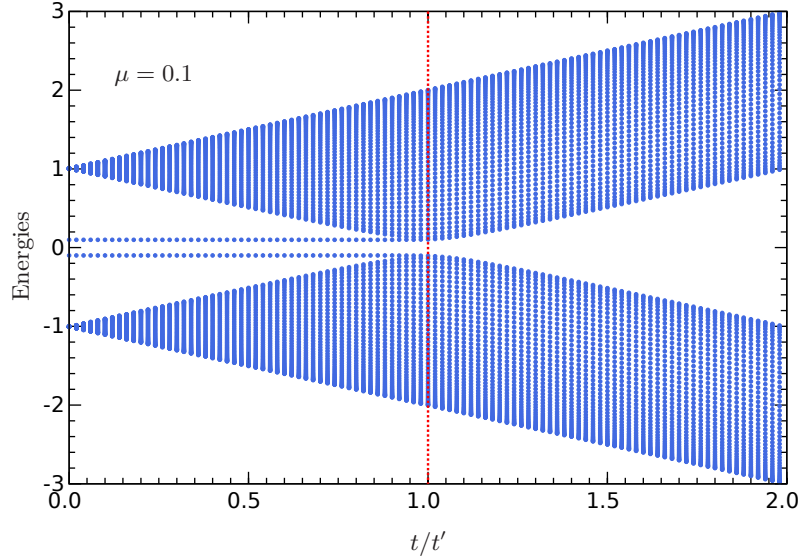}
 \caption{Energy spectrum of SSH model in the presence of staggered potential with
   respect to $t/t'$, for $\mu=0.1$ and $N=100$.
   Non-zero energy edge states appear at $\pm 0.1$,
   when $t/t'<1$.}
\label{energy-ssh}
\end{center}
\end{figure}

Energy spectrum of the SSH model in the presence of staggered potential
has been drawn with respect to $t/t'$, for $\mu=0.1$ and $N=100$
in Figure \ref{energy-ssh}. It indicates that pair of nonzero energy
edge modes appear in the region $0<t/t'<1$. In the absence of
staggered potential, this particular region is topological nontrivial,
when it exhibits a pair of zero-energy edge modes.
Thus it seems that the double degeneracy of edge modes has been lifted
due the presence of staggered field as they appear now at $\pm \mu$. 
Band gap, $\Delta E=2\mu$, exists throughout the region which indicates
no topological phase transition takes place at the point
$t/t'=1$, in the presence of staggered potential. So, the system
always remain in the trivial insulating phase. 

\begin{figure}[h]
  \psfrag{a}{(a)}
  \psfrag{b}{(b)}
  \psfrag{m1}{$\mu=0.1$}
  \psfrag{m3}{$\mu=0.3$}
  \psfrag{m5}{$\mu=0.5$}
   \psfrag{m}{$\mu=1.0$}
   \psfrag{t}{$t/t'$}
\psfrag{c}{$\chi_{ f}(\lambda)$}
\includegraphics[width=330pt]{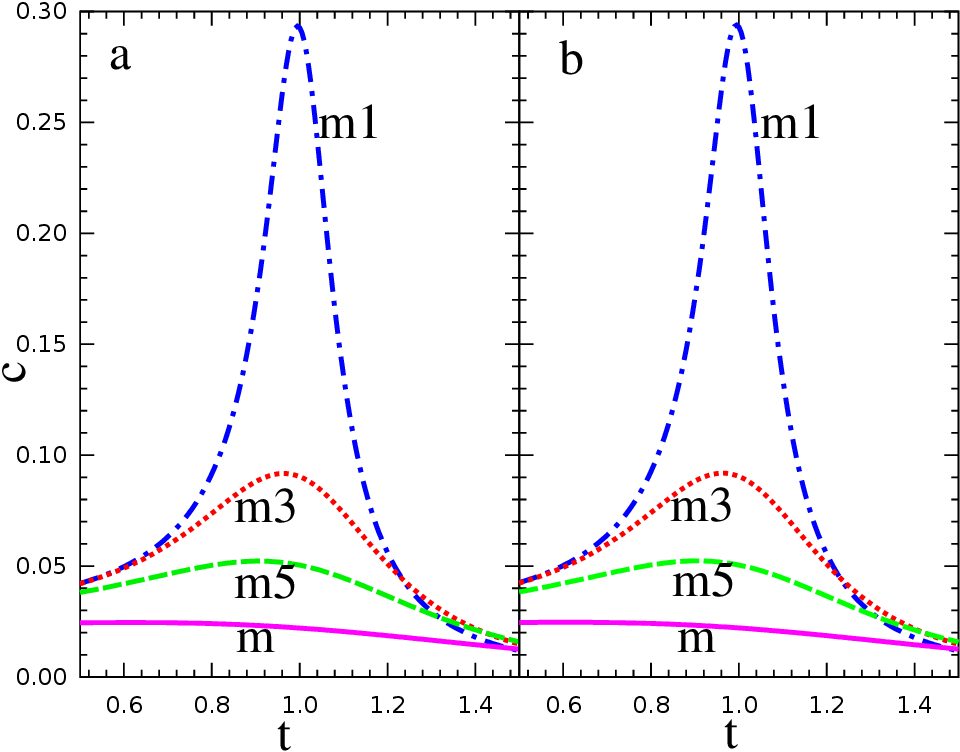}
\caption{Variation of $\chi_{ f}(\lambda)$  with respect to
  $t/t'$  for $N=1001$  in (a)
  and $N=1000$ in (b), for four different values  of
  $\mu=0.1,\,{\rm (blue\,dot\text{-}dashed\, line) },\;
  0.3,\,{\rm (red\, dotted\, line) },\;0.5,\,{\rm (green \,dashed\, line)},\,
  {\rm and}\,1.0$ 
(pink line).}
 \label{chi-even-odd-staggered-potential}
\end{figure}

Behavior of fidelity susceptibility of the SSH model in the presence of
staggered potential has been studied numerically for the open chains of
both odd-$N$ and even-$N$ cases. 
Variation of $\chi_{ f}(\lambda)$ with respect to  $t/t'$
has been shown for four different values of
$\mu=0.1,\,{\rm (blue\,dot\text{-}dashed\, line) },\;
0.3,\,{\rm (red\, dotted\, line) },\;0.5,\,{\rm (green \,dashed\, line)},\,1.0$ 
(pink line), in Figure \ref{chi-even-odd-staggered-potential} (a) and (b),
respectively, for $N=1001$ and $N=1000$. 
No difference is found between the results of $\chi_{ f}(\lambda)$
for odd-$N$ and even-$N$ cases in the presence of staggered potential
as observed in Figure \ref{chi-even-odd-staggered-potential} (a) and (b).
Whereas a small deviation is noted in $\chi_{ f}(\lambda)$ for
odd-$N$ and even-$N$ cases in the absence of staggered potential
as found in Figures \ref{chi-even-odd}. 
Anyway, the remarkable fact is that height of the peak in
$\chi_{ f}(\lambda)$ decreases with the increase of $\mu$, where 
$\chi_{ f}(\lambda)$ for $\mu=1.0$ is almost flat. 
It means no phase transition occurs at the
point $t/t'=1$, even though the average Zak number exhibits
a sudden jump at that point as shown in Figure \ref{zak-ssh}.

Properties of fidelity susceptibility of the SSH model
and the effect of staggered potential on $\chi_{ f}(\lambda)$
have been discussed in this chapter. For this purpose,
numerical results for odd-$N$ and even-$N$ cases have been
presented under the open boundary
condition along with analytic results under periodic boundary condition
only when $\mu=0$. 
The effect of uniform chemical potential will be
trivial in a sense that no change in symmetry will take place
in this case which can be understood easily.
Hamiltonian in the momentum space can be written as
 \be H(k)= \mu\, I+ h_x( k)\, \sigma_x+h_y( k)\, \sigma_y,\label{SSH+mu}
  \ee
where $\mu$ is the potential and $I$ is the $2\times 2$ identity matrix. 
This Hamiltonian (Eq. \ref{SSH+mu}) preserves the same set of
symmetries of the standard SSH model as defined in Eq. \ref{SSH-k}.
The energy spectrum will be shifted upward
by the energy $\mu$, while the eigenvectors remains unchanged
for any value of $\mu$. As a result, no change in topological properties 
will be encountered in the presence of uniform chemical potential.
However, pair of edge modes now appear at energy $\mu$ in the
nontrivial phase. 
In the next section, effect of further neighbor
hopping term on $\chi_{ f}(\lambda)$ will be investigated.
\section{SSH model with further neighbour hopping term}
\begin{figure}[h] 
\psfrag{A}{\large A}
\psfrag{B}{\large B}
\psfrag{v}{\large $t$}
\psfrag{w}{\large $t'$}
\psfrag{z}{\large $t_0$}
\includegraphics[width=300pt]{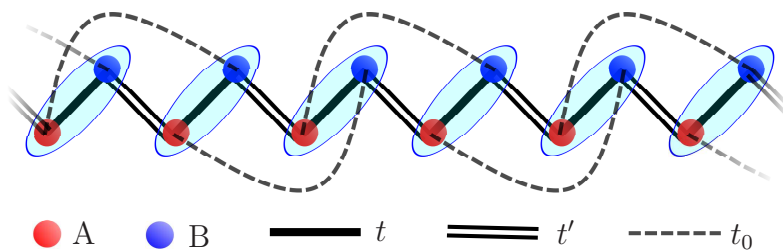}
\caption{SSH model with FN hopping term
  introduced by Li and Miroshnichenko \cite{Li}.}
\label{SSH-1}
\end{figure}

Hamiltonian for SSH model with further neighbour (FN) hopping term
is given by 
\bea
H_{\rm SSH\text{-}FN}&=& H_{\rm SSH}+H_{\rm FN},\\ [0.4em]
H_{\rm FN}&=&\sum_{j=1}^Nt_0\,a^\dag_{j}b_{j+1}+{\rm h.c.},\nonumber
\label{SSH-FN}
\eea
where $t_0$ be the amplitude of the additional FN hopping
term. 
In 2019, Li and Miroshnichenko \cite{Li} showed that a new topological
phase with $\nu=-1$ appears upon introducing
additional terms which allow hopping between sites of A
sublattice and nonadjacent sites of B sublattice but
only among the adjacent unit cells as displayed in Figure \ref{SSH-1}.
It reveals that additional hopping terms are allowed between
the nearest neighbour (NN) unit cells. However,
similar type of hopping between sites of B sublattice and
nonadjacent sites of A sublattice is not permitted. It means only one
further neighbour hopping term per unit cell is introduced within the
adjacent unit cells. 
With this particular choice of sites between which the additional hopping
is taken into account, the resulting system is found to preserve the chiral
symmetry.
 Hamiltonian in the momentum space can be
 expressed as: 
 \be H(k)=\mathbf h'' \cdot {\boldsymbol \sigma},\nonumber
  \ee
  where, 
         \[\left\{\begin{array}{l}h''_x( k)=t+(t'+t_0)\,\cos{( k)}, \\[0.3em]
      h''_y( k)=(t'-t_0)\,\sin{( k)}, \\[0.3em]
       h''_z( k)=0.
         \end{array}\right. \]
Topological phase of the SSH model with FN hopping parameter
  as defined in Eq. \ref{SSH-FN},  in the parameter space
can be described in terms of winding number as \cite{Li,Rakesh1}: 
\be
\nu=\left\{\begin{array}{ll}
    0,& |t'+t_0|<t,\\[0.3em]
    1,&|t'+t_0|>t, \;{\rm and},\;t'>t_0,\\[0.3em]
    -1,&|t'+t_0|>t, \;{\rm and},\;t'<t_0.
  \end{array}\right.
\ee

  \begin{figure}[h]
  \psfrag{a}{(a)}
  \psfrag{b}{(b)}
  \psfrag{c}{(c)} 
  \psfrag{t0}{$t_0$}
    \psfrag{t}{$t=1,\,t'=1$}
   \psfrag{w}{$\nu$}
   \psfrag{E}{Energy}
   \psfrag{F}{$\chi_{f}(t_0)$}
\includegraphics[width=300pt]{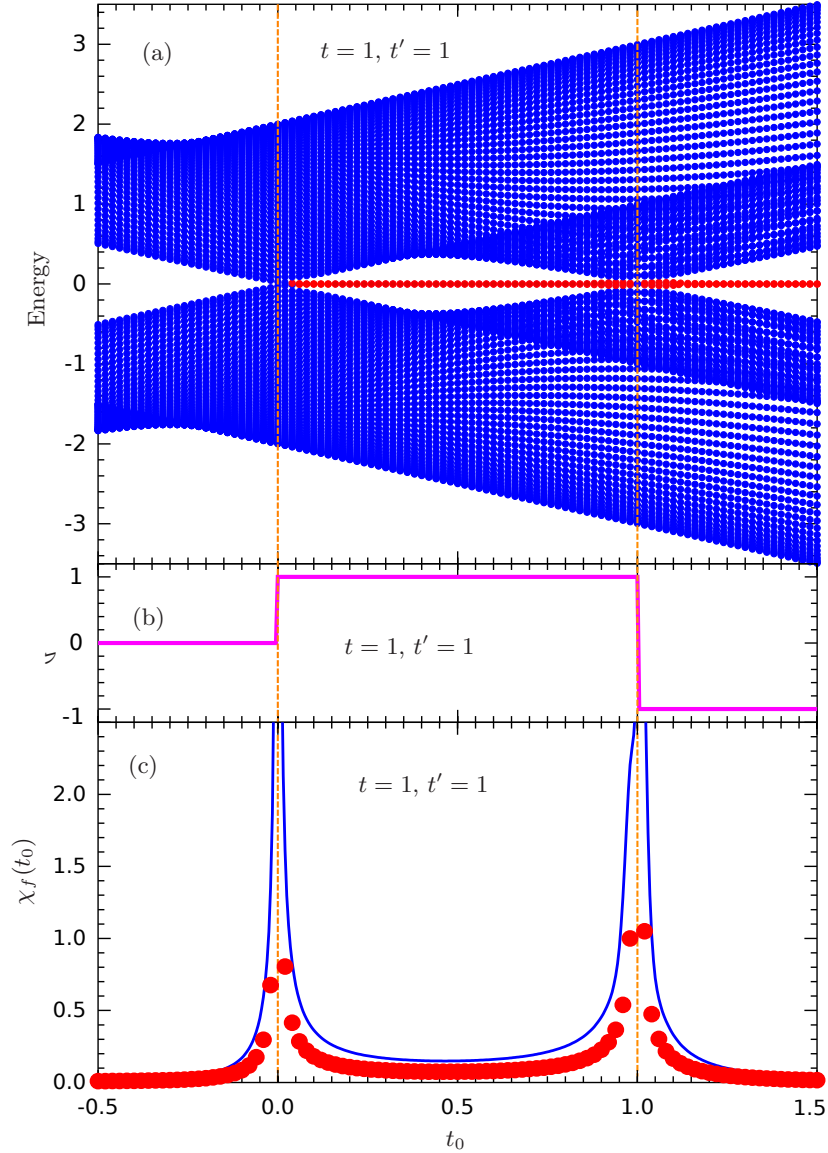}
\caption{Energy spectrum with respect to  $t_0$, in (a), where
  red spheres represent the zero-energy edge sates.
  Variation of winding number with respect to  $t_0$ is shown in pink line (b).
  Variation of $\chi_{f}(t_0)$ has been shown in blue line
  (Eq. \ref{chi-FN}), along with exact
  diagonalization results (Eq. \ref{chi-N}),
  for $N=1000$ in red spheres (c) with respect to  $t_0$. 
  Vertical dashed lines indicate the phase transition points
  at $t_c=0$, and $t_c=1$. 
All these diagrams are drawn for $t=1,\,t'=1$.}
 \label{energy-winding-susceptibility-ssh-FN}
  \end{figure}
  Properties of this extended SSH model 
  have been shown in Figure
  \ref{energy-winding-susceptibility-ssh-FN}, when
  $t=1,\,t'=1$.
  System exhibits two distinct topological phases
  with $\nu=\pm 1$. For examples, system is trivial ($\nu=0$) when
  $t_0<0$, while it possesses a nontrivial phase with $\nu=1$ when
  $0<t_0<1$, and another topological phase with $\nu=-1$ when
  $t_0>1$ as shown in Figure \ref{energy-winding-susceptibility-ssh-FN} (b).
  Variation of $\nu$ with respect to $t_0$ has been plotted in pink line. 
  The topological phase transition are found at $t_0=t_c$, where $t_c=0,\,1$, 
 which have been marked by vertical dashed lines. Phase transition at
  $t_c=0$ reminds the same observed in standard SSH model, while
  that at $t_c=1$ corresponds to the transition between $\nu=1$ and 
  $\nu=-1$, which appears additionally due to the presence of
  FN hopping terms.
  
   Energy spectrum of the system with respect to  $t_0$
  has been drawn in Figure
  \ref{energy-winding-susceptibility-ssh-FN} (a),
   when $t=1,\,t'=1$. The red spheres
  indicates the zero-energy edge modes in the topological phases.
  In each topological phase a pair of zero-energy modes are
  found which is consistent with the bulk-boundary
  correspondence rule for the chiral symmetric model.
  Obviously, zero-energy edge modes are absent in the trivial phase.
  
Fidelity susceptibility in this case is given by 
\be \chi_{ f}(t,t',t_0)=\frac{1}{8\pi} \int_{0}^{\pi} dk
\frac{\sin^2{(k)}(t+2t'\cos{(k)})^2}
    {\left[t^2+(t'-t_0)^2+2t(t'+t_0)\cos{(k)}+4t't_0\cos^2{(k)}\right]^2}.
    \label{chi-FN}
    \ee
     Variation of $\chi_{ f}(t,t',t_0)$ with respect to  $t_0$,
     has been shown in Figure
     \ref{energy-winding-susceptibility-ssh-FN} (c), with blue line
     when $t=1,\,t'=1$.
     Red spheres in (c) represent the
     exact diagonalization data for $N=1000$, using the Eq. \ref{chi-N}, 
     under periodic boundary condition. Very good agreement between analytic and
     exact-diagonalization results are obtained. 
     $\chi_{ f}$ is found to diverge at the phase transition points,
      $t_c=0$, and $t_c=1$, in both analytic and numerical formulations.
Presence of uniform chemical potential does not lead
any change to the topological properties.
On the other hand, these topological phases do not survive
in the presence of staggered potential which is similar to the
case for standard SSH model as discussed before. 

  \begin{figure}[h]
  \psfrag{a}{(a)}
  \psfrag{b}{(b)}
  \psfrag{c}{(c)}
    \psfrag{d}{(d)} 
  \psfrag{t0}{$t_0$}
    \psfrag{t}{$t=1,\,t'=1$}
     \psfrag{E}{Energy}
   \psfrag{chi}{$\chi_{f}(t_0)$}
\includegraphics[width=400pt]{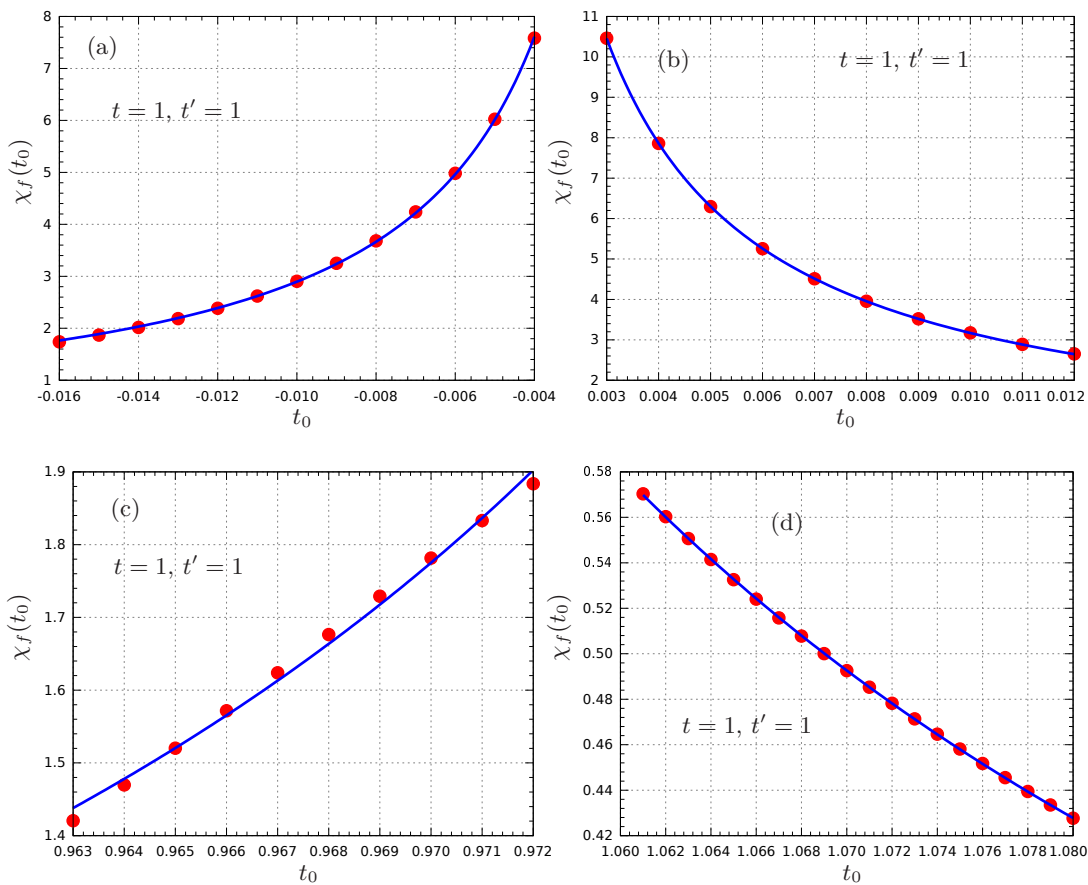}
\caption{Fitting of $\chi_{f}(t_0)$ around the two different
  phase transition points.
  Fitting for left and right sides of  $\chi_{f}(t_0)$ at $t_c=0$
  are shown in (a) and (b), respectively, while that for
  $t_c=1$ are shown in (c) and (d) respectively, 
  when $t=1,\,t'=1$. Red spheres are the values of $\chi_{f}(t_0)$,
while blue line represents the fitting curve.}
\label{fitting}
  \end{figure}
  In order to study the nature of divergence of $\chi_{f}(t_0)$,
  in the vicinity of two phase transition points,
  $t_c=0$  and $t_c=1$, $\chi_{f}(t_0)$ have been fitted with the
   algebraic function, 
  \be
  \chi_{f}(t_0)=\frac{A}{|t_0-t_c|^\beta},
\label{scaling}
\ee
where $\beta$ ($A$) is the
scaling exponent (amplitude), when $t=1,\,t'=1$, which is similar
to the Eq. \ref{scaling-SSH}. 
Both the left and right side of the
  diverging curves have been fitted with the function (Eq. \ref{scaling}),
  and the values of $A$ and $\beta$ for each side have been
  determined for both the transition points.
  For examples, $A=0.0225967,\,(0.032990)$ and $\beta=1.05396,\,(0.991321)$
  are the values of left (right) side of the curve for
  $\chi_{f}(t_0)$ when  $t_c=0$.  Red spheres are the values of $\chi_{f}(t_0)$
  (Eq. \ref{chi-FN}),
  while blue line represents the fitting curve, Eq. \ref{scaling}.
  The fittings for left and right sides
  have been shown in Figure \ref{fitting} (a) and (b),
  respectively.
  Similarly, at another transition point,  $t_c=1$,
 values of the respective quantities are 
 $A=0.052452,\,(0.02957)$ and $\beta=1.03431,\,(1.05770)$
 for the left (right) side of the curve.
 The corresponding fittings for left and right sides
  have been shown in Figure \ref{fitting} (c) and (d),
  respectively. Comparing all these results
  it indicates that the exponents for divergence of $\chi_{f}(t_0)$
  at the two transition points are very close to the value
  $\beta=1$, which is the same to that value obtained analytically for the
  SSH model without the FN hopping terms. So, the presence of FN hopping terms
  does not break any fundamental feature of the fidelity susceptibility.
  It is expected that similar type of investigation
  on the other extended SSH models will reveal the
  same feature at the phase transition points \cite{Rakesh1,Rittwik1,Rakesh2}.
\section{Discussion}
\label{Discussion}
In this chapter, property of fidelity susceptibility of SSH model
under various conditions has been studied.  Fidelity susceptibility
emerges as a very important tool to identify the location
of any kind of phase transition points for many-body system at zero temperature.
Here the location of topological phase transition points
of SSH model and extended SSH model with a single further neighbour
hopping term has been identified with the help of
fidelity susceptibility by means of the point where it diverges.
Expression for fidelity susceptibility has been obtained analytically as well as
its value has been estimated numerically by exact diagonalization method.
In both approaches fidelity susceptibility exhibits sharp peaks at the
phase transition points. Scaling property of the divergence has been
studied and the value of scaling exponent has been obtained. 
For the SSH model both the numeric and analytic results are found to converge,
but, slight deviation is noted in the numeric values of open SSH chains for
odd and even number of sites. This inequality attributes to the
nonequivalent edge modes appear in the different open SSH chains.
Property of fidelity susceptibility in the presence of
staggered and uniform chemical potentials has been explored.
In the same way, nature of fidelity susceptibility for the SSH model with
an additional further neighbour hopping term within adjacent unit cell
has been investigated. In this case two different phase transition points have been
identified.

\section{ACKNOWLEDGMENTS}
AKG gratefully acknowledges Arindam Bhunia for his critical review of this manuscript.
   \section{Data availability statement}
  All data that support the findings of this study are
  included within the article.
   \section{Conflict of interest}
  Authors declare that they have no conflict of interest.

\end{document}